\documentclass[twocolumn,showpacs,preprintnumbers,amsmath,amssymb]{revtex4}

\usepackage{graphicx}
\usepackage{dcolumn}
\usepackage{bm}

\begin{document}

\preprint{APS/123-QED}

\title{Model for the magnetoresistance and Hall coefficient of inhomogeneous graphene}


\author{Rakesh P. Tiwari and D. Stroud}
\affiliation{%
Department of Physics, Ohio State University, Columbus, OH 43210\\
}%

\date{\today}

\begin{abstract}

We show that when bulk graphene breaks into n-type and p-type puddles, the in-plane resistivity becomes strongly field dependent in the presence of a perpendicular magnetic field, even if homogeneous graphene has a field-independent
resistivity.  We calculate the longitudinal resistivity $\rho_{xx}$ and Hall resistivity $\rho_{xy}$
as a function of field for this system, using the effective-medium approximation.  The conductivity tensors
of the individual puddles are calculated using a Boltzmann approach suitable for the band structure of
graphene near the Dirac points.  The resulting resistivity agrees well with experiment, provided
that the relaxation time is weakly field-dependent.  The calculated Hall resistivity has the sign of the
majority carrier and vanishes when there are equal number of n and p type puddles.

\end{abstract}

\maketitle

Graphene is a two-dimensional form of carbon with a hexagonal crystal structure like that of a single
layer of graphite.  Because of this structure, it has the band structure of a semimetal: the Fermi
energy $E_F$ of neutral graphene lies at a ``Dirac point,'' where the electronic density of states $n(E_F) = 0$.  There are two inequivalent Dirac points located at different Bloch vectors ${\bf k}_0$ and ${\bf k}_1$.  Near the Dirac points, the bands are linear functions of the components
of ${\bf k} - {\bf k}_0$ and ${\bf k} - {\bf k}_1$, and $n(E)$ is proportional to $|E - E_F|$.
Because of this unusual band structure, the quasiparticle Hamiltonian near the Dirac points
is formally identical to that of massless Dirac fermions, a feature which is responsible for part
of the recent interest in graphene.

Graphene also has striking transport properties.  For example,
experiments have observed finite conductivity for all values of $E_F$, whether above or below the Dirac point\cite{novoselov}, with a minimum conductivity typically $\sim 4e^2/h$.  However, some workers have suggested that this minimum could have much smaller \cite{miao} or larger \cite{chen} values than $4e^2/h$.  It has been proposed that the existence of a finite conductivity even at the charge neutrality point might be a result of local potential fluctuations, which could cause
a homogeneous neutral graphene sheet to break up into ``puddles'' of electron-rich (n-type) and hole-rich (p-type) character\cite{dassarma}.  These puddles have, in fact, been
unambiguously observed in experiments using scanning tunneling microscopy\cite{martin}. 

Recently measurements of the magnetic-field-dependent longitudinal and Hall resistivity $\rho_{xx}$ and $\rho_{xy}$ measurements have been reported\cite{cho}.  $\rho_{xx}$ was found to increase by nearly tenfold
with increasing magnetic field perpendicular to the graphene film, followed by an apparent saturation at sufficiently strong magnetic field.  These authors found that the magnetoresistance was inconsistent with a two-fluid model of transport by n-type and p-type charge carriers in a homogeneous sheet of graphene, and suggested that it might agree better with a model describing the film as a mixture of n-type and p-type puddles. However, they were able to obtain close agreement between experiment and theory only by assuming an ad hoc empirical form for the magnetoresistance. 

In this Rapid Communication,
we present a simple model for the magnetoresistance and Hall coefficient of
graphene, based on the effective-medium approximation (EMA) in a transverse magnetic field.  Such a model
is reasonable if the n-type and p-type puddles are distributed randomly, as appears to be
the case in Ref.\ \cite{cho}.  Our results show that when the area fractions of $n$-type and $p$-type puddles are exactly equal, $\rho_{xx}$ varies exactly linearly with field.  At other
puddle fractions, it saturates, in 
agreement with experiment.  We find that we can obtain excellent agreement with the observed behavior
of $\rho_{xx}(B)$ if we assume an n-type area fraction $f_n$ satisfying $|f_n-1/2| \sim 0.07$ and 
reasonable values for average carrier density and transport relaxation time.
We also make predictions about the Hall resistivity $\rho_{xy}(B)$. 

We consider magnetotransport in a single layer of graphene subject to a magnetic field ${\bf B} = B\hat{z}$ perpendicular to the graphene layer.  We assume that, because of
a random potential due to charges in the substrate or some other cause, the graphene layer has broken up into a mixture of n-type and p-type puddles, having area fractions $f_n$
and $f_p = 1-f_n$.  We also assume that each of the puddles is large enough to be described by its own magnetoconductivity tensor $\sigma_n$ or $\sigma_p$.  In practice, this assumption means that the puddle dimensions are
larger than a typical carrier mean free path.  This condition may not always be satisfied in practice. in which case the results below might need to be modified.

We assume that $\sigma_n$ and $\sigma_p$ are both given by the usual free-electron (or free-hole) forms, suitably
modified to account for the linear dispersion relations of the electrons and holes near the Dirac point.  Thus,
for the $2 \times 2$ conductivity tensor in the xy plane we write
\begin{eqnarray}
\sigma_n &=& \sigma_0\left[\begin{array}{cc}
\frac{1}{1+(\omega_c\tau)^2} & \frac{\omega_c\tau}{1+(\omega_c\tau)^2} \\
-\frac{\omega_c\tau}{1+(\omega_c\tau)^2} & \frac{1}{1+(\omega_c\tau)^2}
\end{array}\right]
\label{eq:1}
\end{eqnarray}
and
\begin{eqnarray}
\sigma_p &=& \sigma_0\left[\begin{array}{cc}
\frac{1}{1+(\omega_c\tau)^2} & -\frac{\omega_c\tau}{1+(\omega_c\tau)^2} \\
\frac{\omega_c\tau}{1+(\omega_c\tau)^2} & \frac{1}{1+(\omega_c\tau)^2}
\end{array}\right].
\label{eq:2}
\end{eqnarray}
We denote the charge carrier densities in the n-type and p-type puddles by $n$ and $p$, and the corresponding
relaxation times by $\tau_n$ and $\tau_p$.  We also assume that all the puddles have the same density
of charge carriers, so that $n = p$, and that the relaxation times $\tau_n = \tau_p
\equiv \tau$.  With these assumptions, the zero-field conductivities $\sigma_0$ of the n-type and p-type puddles
are equal.  We can also define zero-field mobilities $\mu_n$ and $\mu_p$ by $\sigma_n = ne\mu_n$ and 
$\sigma_p = pe\mu_p$, where $e$ is the magnitude of the electronic charge; with the above assumptions, these mobilities are also equal. 
 
Both the zero-field conductivity $\sigma_0$ of the puddles and the cyclotron frequency $\omega_c$ are modified from their usual free-electron values because of the linear dispersion relations near the Dirac point.  The result
for $\sigma_0$ at temperature $T = 0$ (fully degenerate limit) is 
\begin{equation}
\sigma_0 = \frac{2e^2}{h}v_F\tau\sqrt{\pi n},
\end{equation}
where $v_F$ is the Fermi velocity.  This form is obtained from the usual
solution of the Boltzmann equation for a degenerate Fermi 
gas\cite{ashmer}, which gives 
\begin{equation}
\sigma_{xx} = [2e^2\tau/(2\pi^2)]\int d^2k^\prime\hbar^{-2}
\left(\partial E({\bf k^\prime})/\partial k_x^\prime \right)^2\delta\left(E({\bf k}^\prime) - E_F\right)
\label{eq:cond}
\end{equation}
for the conductivity.  Here, ${\bf k}^\prime$ is the
two-dimensional wave vector, measured relative to one of the Dirac points, and
$E({\bf k}^\prime) = \hbar v_F|{\bf k}^\prime|$ is the energy relative to the Dirac point.  We use
$k_F = \sqrt{\pi n}$, which takes into account the two valleys near the
two inequivalent Dirac points in the graphene band structure, and we have included an extra factor of 2 in
eq.\ (\ref{eq:cond}) for the same reason.
The cyclotron frequency $\omega_c$ is readily
obtained from the semiclassical equation of motion $\hbar\dot{\bf k} = e{\bf v}_{\bf k}\times {\bf B}$, where ${\bf v}_{{\bf k}^\prime} = \hbar^{-1}{\bf \nabla}_{{\bf k}^\prime}E({\bf k}^\prime)$, as applied to a band with the
dispersion relation $E({\bf k}^\prime) = v_F\hbar|{\bf k}^\prime|$; the result is (in SI units),
\begin{equation}
\omega_c = \frac{v_F e B}{\hbar\sqrt{\pi n}},
\end{equation}

Next, we calculate the effective conductivity tensor $\sigma_e$ of a graphene sheet which
has broken up into n-type and p-type puddles.  If $f_n = 1/2$, this would correspond to
the case where the net charge carrier density is zero, corresponding to a neutral graphene sheet which would,
if homogeneous, have its Fermi energy at the Dirac point.  However, it is also possible to have a graphene sheet with $f_n \neq 1/2$, corresponding to a net doping.
This would correspond to a graphene sheet biased by a suitable gate voltage.

A reasonable way of calculating $\sigma_e$ for tensor conductivities is provided by  
the effective-medium approximation (EMA)\cite{stroud}.  In this approach, the
electric fields and currents within the inhomogeneous graphene sheet are calculated as if the
n-type and p-type puddles are compact and approximately circular, and are embedded in an effective medium whose conductivity is calculated self consistently\cite{stroud,guttalstroud}.  The EMA is, in fact, exact at
f = 1/2, provided $\mu_p = \mu_n$\cite{guttalstroud}.  For tensor conductivities, the defining equation for the EMA is
\begin{equation}
\sum_{i=\{n,p\}}f_i\delta\sigma_i\left(I-\Gamma\delta\sigma_i\right)^{-1}=0.
\end{equation}
Here $\delta\sigma_i=\sigma_i-\sigma_{e}$, $I$ is the $2 \times 2$ unit matrix, and, for the
planar geometry considered, $\Gamma=-I/(2\sigma_{e,xx})$ is the depolarization tensor.  This matrix equation reduces to two coupled scalar algebraic equations for
the two independent components of $\sigma_e$ ($\sigma_{e,xx}$ and $\sigma_{e,xy}$), which are easily solved numerically.  The other two components
are determined by $\sigma_{e,yy}=\sigma_{e,xx}$ and $\sigma_{e,yx}=-\sigma_{e,xy}$.  The 
resistivity tensor is then obtained by inverting the matrix $\sigma_{e}$, so that $\rho_{e,xx}=\rho_{e,yy}=\sigma_{e,xx}/(\sigma_{e,xx}^2+\sigma_{e,xy}^2)$ and 
$\rho_{e,xy}=-\rho_{e,yx}=-\sigma_{e,xy}/(\sigma_{e,xx}^2+\sigma_{e,xy}^2)$. 

In order to compare this model to experiment\cite{cho}, we  
need the values of $v_F$, $n$, $\tau$ (or equivalently, $\mu$), and $f_n$.  From the band structure of graphene, $v_F \sim 10^6$ m/sec\cite{novoselov}.  In fact, a value of $1.1 \times 10^6$ m/sec has been inferred from
measurements of the Landau level splitting in graphene\cite{jiang}, and we use this value in the calculations
below.  Also, the measured value of the zero-field resistivity is $\sigma_0^{-1} = \rho_0 = 0.125 h/e^2$.
Given this value, eq.\ (3) provides one condition satisfied by the two parameters $n$ and $\tau$.
We then choose $n$, $\tau$, and $f_n$ so as to best fit the measured $\rho_{xx}(B)$ at $B = 8 T$, and to yield
$\omega_c\tau = 3.1 B$, where B is the magnetic field in T, as reported in Ref.\ \cite{cho}.  
This procedure gives $n \sim 6 \times 10^{14}$ m$^{-2}$ and $f_n \sim 0.43$.  
We find that the best agreement with the resistivity is given at high fields by $\omega_c\tau \sim$ 2.3 B, and at low fields by $\omega_c\tau \sim$ 3.1 B, indicating a weakly field-dependent $\tau$.  The value of $n$ is close to measured value quoted in Ref.\ \cite{cho}.

The calculated results for $\rho_{xx}(B)$ are shown in Fig.\ 1, using these parameters.  As can be seen, the fit to the experimental data is excellent over most of the field range, using
$\omega_c\tau \sim$ 2.3B, and at low fields using $\omega_c\tau \sim$ 3.1B.  The fit, especially at high fields,
is also superior to the two-fluid model discussed (and found inadequate) in Ref.\ \cite{cho}.  
The fit to this puddle model would be nearly perfect over the entire range of
B studied experimentally, if $\tau$ varied by about 30\% as a function of $B$.
The results for $\rho_{xx}$ are independent of the sign of the charge and are thus unchanged
if $f_n \rightarrow 1 - f_n$.

In Fig.\ 2, we show the corresponding results for $\rho_{xy}(B)$.  In this case, we use a field-independent
$\tau$ (corresponding to $\omega_c\tau = 2.3$ B).  We
show results for $f_n = 0.43$, $0.57$, and $0.5$.   $\rho_{xy}(B)$ for $f_n = 0.43$ is equal and opposite to
that for $f_n = 0.57$.  In both cases, $\rho_{xy}$ varies roughly linearly with $B$ for $B$ greater than about 1T.  
At $f_n = 0.5$, $\rho_{xy} = 0$ for all B.  Within the present model, this lattter result is exact, and
not restricted to the EMA\cite{guttalstroud}.  

Fig.\ 3 shows $\rho_{xx}(B, f_n)$ versus$f_n$ for 
several values of B. We use the EMA and the same parameters as in Figs.\ 1 and 2 (with $\omega_c\tau = 2.3$ B). 
$\rho_{xx}$ saturates for all values of $f_n$ except $f_n = 1/2$, for which it increases linearly with $B$.   
Once again, this linearity is exact, and not restricted to the EMA\cite{guttalstroud}.
In Fig.\ 4, we show $\rho_{xy}(B, f_n)$ versus $f_n$ for several values of $B$.  As can
be seen, $\rho_{xy}$ changes sign at $f_n = 1/2$, and approaches a constant as $f_n$ approaches either
1 or 0.  The magnitude of the slope $[d\rho_{xy}/df_n]_{f_n = 1/2}$ increases with increasing $B$, so that, at  large $B$, the Hall resistivity is very close to that of the majority charge carrier. 

The present model agrees well with the measured values of $\rho_{xx}(B)$ in graphene.  However, it is based
on certain assumptions whose validity for graphene we now discuss.
One assumption is that graphene can be treated as a macroscopically inhomogeneous assembly of puddles, each with its own conductivity tensor.  The scanning tunneling microscope images 
shown in Ref.\ \cite{martin} suggest that the carrier density varies appreciably over a distance of perhaps $0.2 \mu$. Using the above density estimates, the number of charge carriers in a puddle of linear dimension $0.2 \mu$ would be $\sim 25$.  This size is rather small to be  treated macroscopically.  On the
other hand, a more reasonable definition of a ``puddle'' might be a region where the charge carriers were all of one sign.  Judging from the images, a typical linear dimension of such a region would be larger than $0.2 \mu$ - perhaps 0.5 - 1$\mu$, and would contain $\sim 500$ carriers. This is probably large 
enough to describe each puddle by its own macroscopic conductivity, provided that the mean free path
$\Lambda$ is less than 1$\mu$.  Even if $\Lambda$ were larger than this, one could
still use the semiclassical model with a size-limited $\Lambda$ of magnitude $\Lambda
\sim a$, where $a$ is the linear dimension of the puddle.  This would give $\tau \sim a/v_F \sim 4.5 \times 10^{-13}$
sec.  In short, treatment of the puddle mixture macroscopically is probably appropriate in the
case of some disordered samples of graphene, and the results seem to agree with experiment.

Another point, as can be seen from Fig.\ 3, is that the $\rho_{xx}(B, f_n)$ 
saturates (approaches a finite limit at large $B$) only if $f_n \neq 1/2$, i.\ e., if there is a net charge imbalance produced by a suitable gate voltage. Such saturation seems to be observed in experiments\cite{cho}.
But, as can be seen from Figs.\ 2 and 4, a charge imbalance would lead to a nonzero $\rho_{xy}(B)$.  It would be of interest if $\rho_{xy}(B)$ could be measured and compared to the values needed for the present model to agree with
experiment. 

Thirdly, the present model treats the electron dynamics semiclassically, and thus does not take into account
the quantum Hall effect (QHE), which is seen at sufficiently high fields\cite{qhe,jiang}.
Typically, the QHE will become visible when the spacing between the Landau levels is large compared to $k_BT$.  This can occur even at room temperature
in graphene\cite{qhe}. 
 If the QHE becomes important, the present semiclassical model would need to be modified.  

Finally, we obtain the best fits to experiment if we assume a weakly magnetic-field dependent relaxation time,
as described above.  Such field-dependence could be reasonable, but it would be useful to have a model which explicitly produces a magnetic field-dependent $\tau$.

We find that our results are quite insensitive to slight changes in the parameters or other features of the model.  For example Fig.\ 1 suggests that $\rho_{xx}$ changes only slightly, but not dramatically, when $\tau$ is varied by $\sim 30$\%.  Also, we have recalculated $\rho_{e,xx}(B,T)$ without the assumption that the electrons and holes have equal mobilities.  Even if the mobilities are different, we find that $\rho_{xx}(B)$ still varies linearly with $B$ at $f_n = 0.5$ and saturates at other values of $f$.  Another change in our model is suggested by that fact that
the carrier density in graphene must be a continuous function of position, rather than being simply bimodal as
postulated in our model.  To check the effects of a non-bimodal distribution, we have repeated our calculations assuming {\it four} types of puddles, two n-type and two p-type, with two different densities each of electrons
and holes.  Once again, we find that the resulting $\rho_{xx}(B,f_n)$ depends primarily on $f_n$ and $\tau$, and not on
the presence of two types of n and of p puddles.  Finally, we have considered the case of a three-component
composite, made up of n-type, p-type, and insulating regions.  Here, once again, we find that, for a small insulating areal faction ($\sim 0.1$), we obtain linear magnetoresistance if $f_n = f_p$ and saturating magnetoresistance otherwise, similar to the case of no insulating regions.  Thus, our results are not much affected by small modifications in our model.

To summarize, we have calculated the magnetoresistance and Hall resistivity for a semiclassical model of
graphene, on the assumption that it is a mixture of n-type and p-type puddles, and using the correct form
of the band structure near the Dirac points.  The resulting magnetoresistance is in good agreement
with experiment, provided that the areal fractions of n and p-type puddles are slightly different and that the
relaxation time is weakly magnetic-field dependent.  Further confirmation of the model could be obtained if
the measured Hall resistivity were compared to that computed from this model.        
   
Funding for this research was provided by the Center for Emergent Materials at the Ohio State University, an NSF MRSEC (Award No.\ DMR-0820414).

\begin{figure}[h]
\begin{center}
\includegraphics[scale=0.4,angle=270]{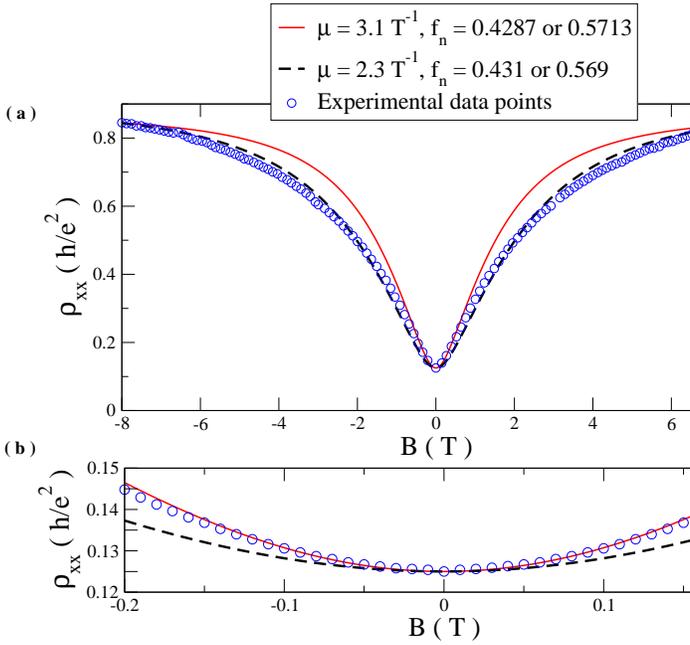}
\caption{(Color online.)  $\rho_{xx}(B, f_n)$ as a function of $B$ (in $T$), with two different assumptions about the
mobility,  In both cases, we assume that the electron and hole mobilities are equal.  Solid (red) line: calculated results, assuming $\mu \equiv \omega_c\tau/B = 3.1 T^{-1}$ and $f_n = 0.429$ (or $0.571$)..  Dashed (black) line: calculated results with $\mu = 2.3 T^{-1}$ and $f_n = 0.431$ (or $0.569$). Open circles are experimental data from Ref.\ \cite{cho}.  Lower panel is a blowup of the calculations and data from the upper panel.}
\end{center}
\label{fig:1}
\end{figure}

\begin{figure}[h]
\begin{center}
\includegraphics[scale=0.4,angle=270]{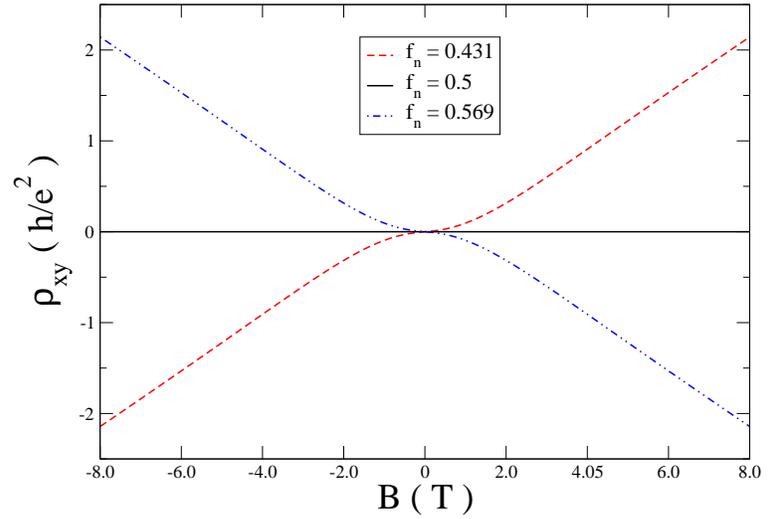}
\caption{(Color online.) Calculated $\rho_{xy}(B, f_n)$, as obtained in the EMA, using $\mu = 2.3T ^{-1}$ and $f_n = 0.431$, $0.569$, and $0.5$.}
\end{center}
\label{fig:2}
\end{figure}

\begin{figure}[h]
\begin{center}
\includegraphics[scale=0.4,angle=270]{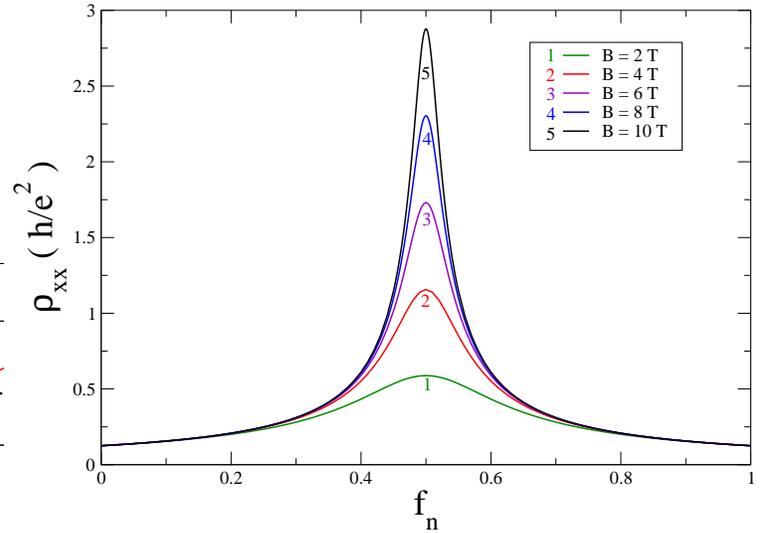}
\caption{(Color online.)  $\rho_{xx}(B, f_n)$ as a function of $f_n$ for several values of $B$ and $\mu = 2.3 T^{-1}$}
\end{center}
\label{fig:3}
\end{figure}

\begin{figure}[h]
\begin{center}
\includegraphics[scale=0.4,angle=270]{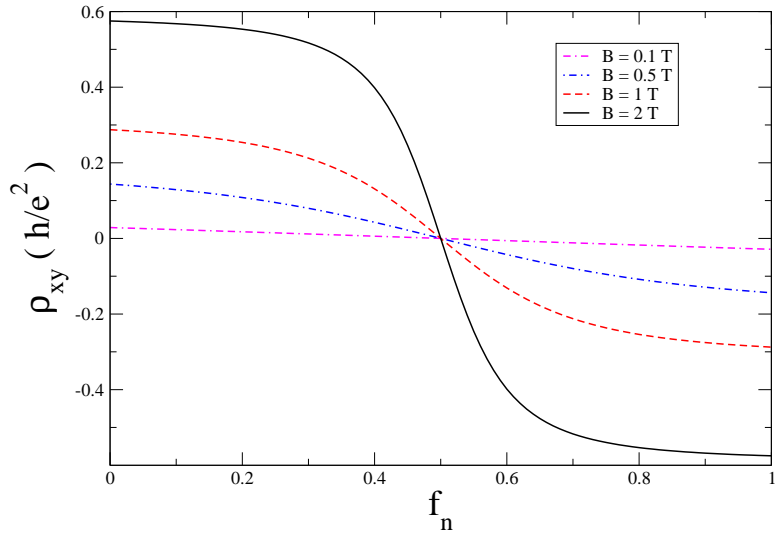}
\caption{(Color online.)  Calculated $\rho_{xy}$ as a function of $f_n$ for several different values of $B$, using
$\mu = 2.3 T^{-1}$.}
\end{center}
\label{fig:4}
\end{figure}


\newpage



\begin{thebibliography}{99}

\bibitem{novoselov} K. S. Novoselov, A. K. Geim, S. V. Morozov, D. Jiang, M. I. Katsnelson, I. V. Grigorieva, S. V. Dubonos and A. A. Firsov, Nature (London) {\bf 438}, 
197 (2005).
\bibitem{miao} F. Miao, S. Wijeratne, Y. Zhang, U. C. Coskun, W. Bao and C. N. Lau, Science {\bf 317}, 1530 (2007).
\bibitem{chen} J. H. Chen, M. Ishigami, C. Jang, D. R. Hines, M. S. Fuhrer and E. D. Williams, Adv. Mater. (Weinheim, Ger.) {\bf 19}, 3623 (2007).
\bibitem{dassarma} S. Adam, E. H. Hwang, V. M. Galitski and S. D. Sarma, Proc. Natl. Acad. Sci. U.S.A. {\bf 104}, 18392 (2007).
\bibitem{martin} J. Martin, N. Akerman, G. Ulbricht, T. Lohmann, J. H. Smet, K. Von Klitzing and A. Yacoby, Nat. Phys. {\bf 4}, 148 (2008)
\bibitem{ashmer}  See, e.\ g., N.\ W.\ Ashcroft and N.\ D.\ Mermin, {\it Solid State Physics},
(Saunders, Orlando, FL, 1976), eq.\ (12.42). 
\bibitem{cho} S. Cho and M. S. Fuhrer, Phys. Rev. B. {\bf 77}, 081402(R) (2008).
\bibitem{jiang} Z.\ Jiang, E.\ A.\ Henriksen, L.\ C.\ Tung, Y.-J.\ Wang, M.\ E.\ Schwartz, M.\ Y.\ Han,
P.\ Kim, and H.\ L.\ Stormer, Phys.\ Rev.\ Lett.\ {\bf 98}, 197403 (2007).
\bibitem{stroud} D. Stroud, Phys. Rev. B {\bf 12}, 3368 (1975).
\bibitem{guttalstroud} V. Guttal and D. Stroud, Phys. Rev. B {\bf 71}, 201304(R) (2005).
\bibitem{vf} See, e.\ g., C.\ Berger {\it et al.}, Science {\bf 312}, 1191(2006).
\bibitem{qhe} K. S. Novoselov \textit{et al.}, Science {\bf 315}, 1379 (2007)
\end{thebibliography}
\end{document}